\documentclass[a4paper,11pt]{article}
\usepackage{pos}

\def\tac{\langle \sigma_\text{ann} v\rangle}

\def\deg{^{\circ}}

\title{Search for Dark Matter Annihilation Signals in the Galactic Center Halo with VERITAS}
 \ShortTitle{Search for Dark Matter in the Galactic Center with VERITAS}

\author*[a]{James L. Ryan}

\affiliation[a]{SLAC National Accelerator Laboratory/Kavli Institute for Particle Astrophysics and Cosmology,\\
  2575 Sand Hill Road, Menlo Park, California 94025, USA}

\onbehalf{for the VERITAS collaboration} 

\emailAdd{jlryan@slac.stanford.edu}

\abstract{
Gamma-ray observations of the Galactic Center (GC) region provide some of the most sensitive measurements of annihilation signals from dark matter in the GeV--TeV mass range.
We present the first results from 178 hours of VERITAS observations of a 2 degree region around the GC, taken between 2010 and 2022.
The analysis uses a newly-developed template-based background method, as well as analysis techniques optimized for observations taken at large zenith angles.
No significant excess is found in any region of interest considered.
We derive model-independent limits on the dark matter thermally-averaged annihilation cross section which are among the strongest from any indirect detection experiment for TeV-scale masses.}

\ConferenceLogo{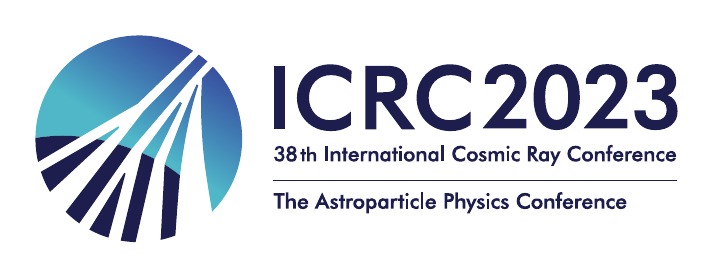}

\FullConference{%
38th International Cosmic Ray Conference (ICRC2023)\\
  26 July - 3 August, 2023\\
  Nagoya, Japan}

\begin{document}
\maketitle

\section{Introduction}
Gamma-ray observations are used to perform some of the most sensitive searches for dark matter \citep{aramaki22}.
In certain models, dark matter particles can annihilate or decay into Standard Model particles which eventually yield gamma rays, among other end states.
An indirect detection of dark matter may be accomplished by detecting these gamma rays.

The strength of a gamma-ray signal due to dark matter is proportional to the number density of dark matter particles along the line of sight, quantified by the J-factor \citep{bergstrom98}.
High J-factors may be found toward the Galactic Center, nearby dwarf spheroidal galaxies (dSphs), and nearby external galaxies.
While the highest J-factor lies toward the Galactic Center, where the Galactic dark matter halo density is highest, the region also contains astrophysical sources of gamma rays.
The region is host to the supermassive black hole Sagittarius A* (Sgr A*), pulsar wind nebulae, supernova remnants, and a sea of cosmic rays and dense molecular gas.
Consequently, specialized analysis techniques are sometimes necessary to study the region.

We present the results of a recent VERITAS analysis of the Galactic Center halo, with implications for dark matter \citep{adams21}.
We describe improvements made to event reconstruction for the Galactic Center data observing conditions, as well as the development of a template-based background estimation technique.
Model-independent limits are derived for the thermally-averaged dark matter annihilation cross section.

\section{VERITAS Data and Analysis}
The Very Energetic Radiation Imaging Telescope Array System (VERITAS) is an imaging atmospheric Cherenkov telescope (IACT) array located at the Fred Lawrence Whipple Observatory in southern Arizona at $31\deg \; 40'$ N, $110\deg \; 57'$ W, 1.3 km above sea level.
VERITAS is sensitive to gamma rays with energies from 100 GeV to over 30 TeV, with an energy resolution of 15--25\% and angular resolution $<0.1\deg$ at 1 TeV.
VERITAS can make a significant detection of a source 1\% the brightness of the Crab in $\sim$25 hours.
Significant upgrades to the VERITAS cameras and trigger system were made in summer 2012 \citep{kieda13}---our data taken before and after this are referred to as V5 and V6.

We have observed Sgr A* for 177.8 hours between 2010 April and 2022 June with VERITAS. 
Additionally, we have taken 50.8 hours of off-source observations of a nearby patch of sky with no known gamma-ray sources and few bright stars, which we refer to as Sgr A* Off.
Observations are taken with the telescope pointing offset by $0.5\deg$ or $0.7\deg$ from the nominal source position,
which is (RA,Dec)=$(266.417\deg,-29.0078\deg)$ for Sgr A* and (RA,Dec)=($260.917\deg,-29.0078\deg)$ for Sgr A* Off, in equatorial coordinates (J2000).
Data runs exhibiting large anomalies in triggering rates, sky temperatures, or other metrics are not included in the analysis.
Short segments of runs showing anomalous behavior (e.g. due to clouds or car headlights) are also excluded.
The total exposure times remaining after quality cuts and deadtime corrections are 154.3 hours for Sgr A* and 44.0 hours for Sgr A* Off.

All observations are taken at large zenith angles (LZA), since Sgr A* does not rise more than $30\deg$ above the horizon at the location of VERITAS.
LZA observations result in higher effective area, but also higher energy threshold.
The standard direction and energy reconstruction algorithms used by VERITAS also degrade at LZA.
To remedy this, a displacement method for directional reconstruction was implemented using boosted decision trees (BDTs).
The BDTs are trained using image length, width, and size, as well as 
the arrival time gradient across the image and the integrated charge lost due to image truncation at the camera edges.
Separate BDTs are trained for different observation zenith angles, azimuth angles, and pedestal noise levels.
BDTs were also trained to reconstruct event energy using the same parameters.
We obtain angular resolution ${\sim}0.12\deg$ and energy resolution ${\sim}19\%$.
Different gamma/hadron separation cuts are also used, optimized using LZA observations of the Crab Nebula.

\subsection{Template Background Method}
Not all cosmic-ray events can be rejected at the gamma/hadron separation step.
A background estimation step is therefore included to quantify the number of cosmic ray events expected to remain among the ``gamma-like'' events.
Typically, regions spatially offset from the region of interest within the same field of view are used, as in the ring background and reflected region methods \citep{berge07}.
However, crowded fields of view and extended sources complicate the implementation of these methods.
We instead model the background using templates, avoiding the use of gamma-like events from the Sgr A* data as inputs.
Background methods using templates have been used in IACT analyses previously \citep{rowell03,mohrmann19}.

\begin{figure}[t]
\centering
\includegraphics[width=\textwidth]{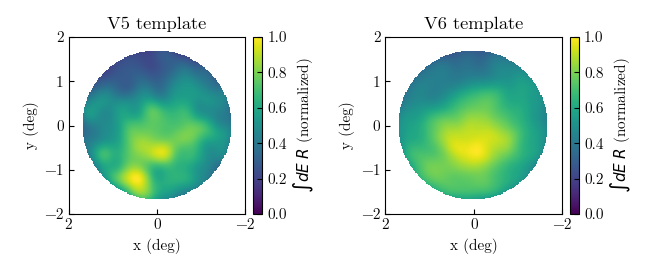}
\caption{
Normalized templates integrated over energy in camera coordinates for observing epochs V5 (left) and V6 (right). The camera axes align with the azimuth and zenith axes, and asymmetries about these axes are visible.
}
\label{fig:templates}
\end{figure}

We construct templates of expected count rates in bins of camera coordinates and energy.
The templates are derived from the Sgr A* Off data, by summing the counts and livetime in each 3D template bin and then calculating rates.
Separate templates are derived for V5 and V6 data, and are shown in Figure \ref{fig:templates}.
The rates decrease with increasing distance from the camera center,
and a gradient is visible in the $y$ direction. 
Additional smaller-scale features can also be seen.

The templates require an additional overall normalization factor to describe individual runs.
Using the number of events determined to be hadron-like during gamma/hadron separation, as well as observation season and zenith angle, the normalization can be predicted with ${\sim}14\%$ accuracy.

For the template background method, the likelihood $L$ of observing $N_\text{on}$ gamma-like events is that of the Gaussian-mean background problem, given by \citep{cousins08} 
\begin{equation} 
L =
\frac{(N_\text{S}+N_\text{B})^{N_\text{on}}}{N_\text{on}!} e^{-(N_\text{S}+N_\text{B})}
\frac{1}{\sqrt{2\pi \sigma_\text{B}^2}}e^{-\frac{1}{2}(N_\text{B}-\mu_\text{B})^2/\sigma_\text{B}^2}
\end{equation}
where $N_\text{S}$ is the number of signal counts and
$N_\text{B}$ is the number of background counts, which follows a Gaussian distribution with mean $\mu_\text{B}$ and standard deviation $\sigma_\text{B}$.
Significance calculations follow from the likelihood.

\begin{figure}[t]
\centering
\includegraphics[width=\textwidth]{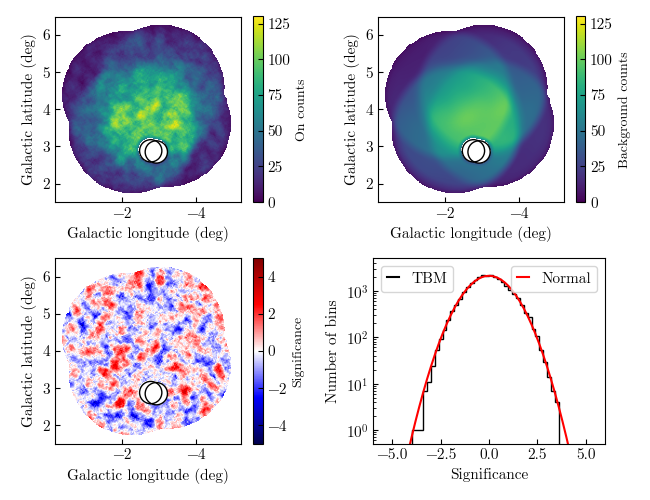}
\caption{
Sgr A* Off analysis using the template background method (TBM). 
Maps are shown of correlated On counts (top left), mean background counts (top right), and significance (bottom left).
Exclusion regions around bright stars d Ophiuchi and HIP85442 (black) are indicated.
A histogram of significances is also shown (bottom right).
The red curve is the normal distribution multiplied by the total number of 2D significance map bins.
}
\label{fig:tbmvalidation}
\end{figure}

An analysis of the Sgr A* Off data using the template background method is shown in Figure \ref{fig:tbmvalidation}.
As the field is free of gamma-ray sources, significances calculated at different positions should be normally distributed.
We calculate significances in circular regions of $0.13\deg$ radius and find them to be normally distributed, while the 2D significance map does not show correlated residuals on scales larger than the chosen size of the region of interest.

\section{Dark Matter Limits}
\begin{figure}[t]
\centering
\includegraphics[width=\textwidth]{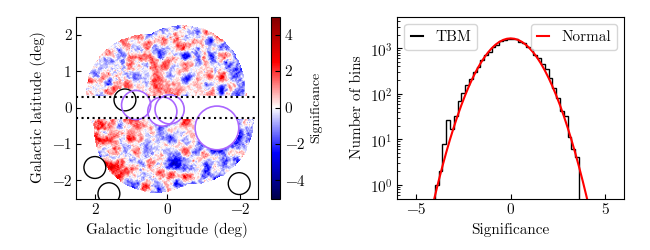}
\caption{
Sgr A* significance map (left) and histogram (right) resulting from the template background method analysis.
Exclusion regions around bright stars (black), gamma-ray sources (purple), and the GC ridge (dotted line) are indicated.
The red curve is the normal distribution multiplied by the total number of 2D significance map bins.
}
\label{fig:sgr_dm}
\end{figure}

The Galactic Center halo is analyzed with the template background method to search for a potential dark matter signal.
We use 15 $0.1\deg$-wide ring-shaped regions of interest centered on Sgr A*, to take advantage of the dark matter halo’s spatial profile, as was done in analyses performed by the H.E.S.S. collaboration \citep{hess16,hess22}.
The rings lie between $0.5\deg$ and $2\deg$ from the Galactic Center.
17 logarithmically-spaced energy bins between 2 TeV and 100 TeV are used.
Bright stars and known gamma-ray sources are excluded from the analysis, including Galactic ridge emission between $\pm0.3\deg$ Galactic latitude, as shown in Figure \ref{fig:sgr_dm}.
No other significant sources of gamma rays are detected.

We use our observations to place an upper limit on the thermally-averaged dark matter annihilation cross section $\tac$.
The the differential flux of gamma rays from annihilating dark matter $d\Phi/dE$ is given by 
\begin{equation} 
\frac{d\Phi}{dE}= \frac{1}{4\pi}\frac{\tac}{2m_\text{DM}^2}\frac{dN}{dE}\int d\Omega\int_\text{los}d s \; \rho_\text{DM}^2
\end{equation}
where $E$ is the gamma ray's energy, $m_\text{DM}$ is the dark matter particle mass, $dN/dE$ is the differential energy spectrum of particles produced per annihilation, $\rho_\text{DM}$ is the dark matter density, $\Omega$ is solid angle, and $s$ is distance along the line of sight.
We describe the dark matter distribution with an Einasto profile with scale radius 20 kpc and $\alpha = 0.17$, normalized to $0.4 \text{GeV} \text{cm}^{-3}$ at a distance of 8.5 kpc from the GC.

\begin{figure}[t]
\centering
\includegraphics[width=\textwidth]{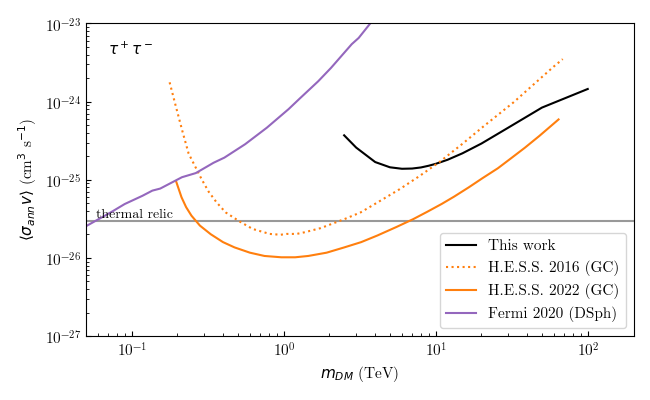}
\caption{
95\% confidence-level upper limits on $\tac$, assuming an Einasto profile (black line) and annihilation into $\tau^+\tau^-$.
Constraints from H.E.S.S. observations of the GC \citep[orange;][]{hess16,hess22} and \textit{Fermi}-LAT observations of DSphs \citep[purple;][]{hoof20} are also shown.
}
\label{fig:limtau}
\end{figure}

Particle model-independent limits are derived in each particle production channel using a profile likelihood method, with the $dN/dE$ spectra in \citep{cirelli11}.
Our constraints in the $\tau^+\tau^-$ channel are shown in Figure \ref{fig:limtau},
where a 95\% confidence level upper limit of $\tac < 1.38\times 10^{-25}\text{ cm}^3\text{ s}^{-1}$ at $m_\text{DM} = 6$ TeV is obtained.
This represents the strongest dark matter limit from VERITAS.
This result, with the slightly stronger H.E.S.S. GC measurement \citep{hess22}, also represent the strongest limits from any measurement to date for dark matter masses between $\sim$2 TeV and 100 TeV.
Our limit approaches that of \citep{hess22} despite the difference in live time (154.3 hr versus 546 hr) due to the effective area increase at LZA, resulting in a similar product of effective area and live time, ${\sim}2\times 10^{11} \text{ m}^2\text{ s}$ above $\sim$10 TeV near the GC.

\section{Conclusions}
We have presented the results of a VERITAS dark matter search using gamma rays from the Galactic Center halo.
A template-based background method is developed that allows for the analysis of fields where gamma-ray emission may be present over the entire field of view.
The Galactic Center dark matter search that will be conducted by Cherenkov Telescope Array will also use templates to model residual cosmic-ray background events \citep{acharyya21}.
We obtain among the strongest limits on the thermally-averaged annihilation cross section for dark matter with masses between 2 and 100 TeV by any experiment.

This research is supported by grants from the U.S. Department of Energy Office of Science, the U.S. National Science Foundation and the Smithsonian Institution, by NSERC in Canada, and by the Helmholtz Association in Germany. This research used resources provided by the Open Science Grid, which is supported by the National Science Foundation and the U.S. Department of Energy's Office of Science, and resources of the National Energy Research Scientific Computing Center (NERSC), a U.S. Department of Energy Office of Science User Facility operated under Contract No. DE-AC02-05CH11231. We acknowledge the excellent work of the technical support staff at the Fred Lawrence Whipple Observatory and at the collaborating institutions in the construction and operation of the instrument.

\clearpage

\section*{Full Author List: VERITAS Collaboration}

\scriptsize
\noindent
A.~Acharyya$^{1}$,
C.~B.~Adams$^{2}$,
A.~Archer$^{3}$,
P.~Bangale$^{4}$,
J.~T.~Bartkoske$^{5}$,
P.~Batista$^{6}$,
W.~Benbow$^{7}$,
J.~L.~Christiansen$^{8}$,
A.~J.~Chromey$^{7}$,
A.~Duerr$^{5}$,
M.~Errando$^{9}$,
Q.~Feng$^{7}$,
G.~M.~Foote$^{4}$,
L.~Fortson$^{10}$,
A.~Furniss$^{11, 12}$,
W.~Hanlon$^{7}$,
O.~Hervet$^{12}$,
C.~E.~Hinrichs$^{7,13}$,
J.~Hoang$^{12}$,
J.~Holder$^{4}$,
Z.~Hughes$^{9}$,
T.~B.~Humensky$^{14,15}$,
W.~Jin$^{1}$,
M.~N.~Johnson$^{12}$,
M.~Kertzman$^{3}$,
M.~Kherlakian$^{6}$,
D.~Kieda$^{5}$,
T.~K.~Kleiner$^{6}$,
N.~Korzoun$^{4}$,
S.~Kumar$^{14}$,
M.~J.~Lang$^{16}$,
M.~Lundy$^{17}$,
G.~Maier$^{6}$,
C.~E~McGrath$^{18}$,
M.~J.~Millard$^{19}$,
C.~L.~Mooney$^{4}$,
P.~Moriarty$^{16}$,
R.~Mukherjee$^{20}$,
S.~O'Brien$^{17,21}$,
R.~A.~Ong$^{22}$,
N.~Park$^{23}$,
C.~Poggemann$^{8}$,
M.~Pohl$^{24,6}$,
E.~Pueschel$^{6}$,
J.~Quinn$^{18}$,
P.~L.~Rabinowitz$^{9}$,
K.~Ragan$^{17}$,
P.~T.~Reynolds$^{25}$,
D.~Ribeiro$^{10}$,
E.~Roache$^{7}$,
J.~L.~Ryan$^{22}$,
I.~Sadeh$^{6}$,
L.~Saha$^{7}$,
M.~Santander$^{1}$,
G.~H.~Sembroski$^{26}$,
R.~Shang$^{20}$,
M.~Splettstoesser$^{12}$,
A.~K.~Talluri$^{10}$,
J.~V.~Tucci$^{27}$,
V.~V.~Vassiliev$^{22}$,
A.~Weinstein$^{28}$,
D.~A.~Williams$^{12}$,
S.~L.~Wong$^{17}$,
and
J.~Woo$^{29}$\\
\\
\noindent
$^{1}${Department of Physics and Astronomy, University of Alabama, Tuscaloosa, AL 35487, USA}

\noindent
$^{2}${Physics Department, Columbia University, New York, NY 10027, USA}

\noindent
$^{3}${Department of Physics and Astronomy, DePauw University, Greencastle, IN 46135-0037, USA}

\noindent
$^{4}${Department of Physics and Astronomy and the Bartol Research Institute, University of Delaware, Newark, DE 19716, USA}

\noindent
$^{5}${Department of Physics and Astronomy, University of Utah, Salt Lake City, UT 84112, USA}

\noindent
$^{6}${DESY, Platanenallee 6, 15738 Zeuthen, Germany}

\noindent
$^{7}${Center for Astrophysics $|$ Harvard \& Smithsonian, Cambridge, MA 02138, USA}

\noindent
$^{8}${Physics Department, California Polytechnic State University, San Luis Obispo, CA 94307, USA}

\noindent
$^{9}${Department of Physics, Washington University, St. Louis, MO 63130, USA}

\noindent
$^{10}${School of Physics and Astronomy, University of Minnesota, Minneapolis, MN 55455, USA}

\noindent
$^{11}${Department of Physics, California State University - East Bay, Hayward, CA 94542, USA}

\noindent
$^{12}${Santa Cruz Institute for Particle Physics and Department of Physics, University of California, Santa Cruz, CA 95064, USA}

\noindent
$^{13}${Department of Physics and Astronomy, Dartmouth College, 6127 Wilder Laboratory, Hanover, NH 03755 USA}

\noindent
$^{14}${Department of Physics, University of Maryland, College Park, MD, USA }

\noindent
$^{15}${NASA GSFC, Greenbelt, MD 20771, USA}

\noindent
$^{16}${School of Natural Sciences, University of Galway, University Road, Galway, H91 TK33, Ireland}

\noindent
$^{17}${Physics Department, McGill University, Montreal, QC H3A 2T8, Canada}

\noindent
$^{18}${School of Physics, University College Dublin, Belfield, Dublin 4, Ireland}

\noindent
$^{19}${Department of Physics and Astronomy, University of Iowa, Van Allen Hall, Iowa City, IA 52242, USA}

\noindent
$^{20}${Department of Physics and Astronomy, Barnard College, Columbia University, NY 10027, USA}

\noindent
$^{21}${ Arthur B. McDonald Canadian Astroparticle Physics Research Institute, 64 Bader Lane, Queen's University, Kingston, ON Canada, K7L 3N6}

\noindent
$^{22}${Department of Physics and Astronomy, University of California, Los Angeles, CA 90095, USA}

\noindent
$^{23}${Department of Physics, Engineering Physics and Astronomy, Queen's University, Kingston, ON K7L 3N6, Canada}

\noindent
$^{24}${Institute of Physics and Astronomy, University of Potsdam, 14476 Potsdam-Golm, Germany}

\noindent
$^{25}${Department of Physical Sciences, Munster Technological University, Bishopstown, Cork, T12 P928, Ireland}

\noindent
$^{26}${Department of Physics and Astronomy, Purdue University, West Lafayette, IN 47907, USA}

\noindent
$^{27}${Department of Physics, Indiana University-Purdue University Indianapolis, Indianapolis, IN 46202, USA}

\noindent
$^{28}${Department of Physics and Astronomy, Iowa State University, Ames, IA 50011, USA}

\noindent
$^{29}${Columbia Astrophysics Laboratory, Columbia University, New York, NY 10027, USA}

\end{document}